\documentclass[conference]{IEEEtran}
\usepackage{blindtext, graphicx}
\usepackage{xcolor}
\usepackage{hyperref}
\usepackage{booktabs}
\usepackage{colortbl}
\usepackage{amsmath}
\makeatletter 
\g@addto@macro\normalsize{%
  \setlength\abovedisplayskip{-2pt}
  \setlength\belowdisplayskip{-1pt}
  \setlength\abovedisplayshortskip{-2pt}
  \setlength\belowdisplayshortskip{-1pt}
}
\makeatother
\usepackage{amssymb}
\usepackage{pgfplots}
\usepackage{tikz}
\usepackage{xcolor}
\usepackage{float}
\usepackage{graphicx}
\usepackage{color}
\usepackage{chemist}
\pgfplotsset{compat=1.6}
\usepackage{multicol,lipsum}
\usepackage{multirow}
\usepackage{changepage}
\usepackage{eurosym}
\usepackage{hyperref}
\usepackage[justification=centering]{caption}
\usepackage{relsize}
\usepackage{wasysym}
\usepackage{tabularx}
\usepackage{wrapfig}
\usepackage{booktabs}
\usepackage{caption}
\usepackage[numbers,sort&compress]{natbib}
\usepackage{titlesec}
\usepackage[capitalise]{cleveref}

\usepackage[nodisplayskipstretch]{setspace}
\ifCLASSINFOpdf
\else
\fi
\begin{document}

%
\title{The Role of Extended Horizon Methodology in Renewable-Dense Grids With Inter-Day Long-Duration Energy Storage\vspace{-3ex}}


\author{
    \IEEEauthorblockN{
        Amogh A. Thatte\IEEEauthorrefmark{1}, Sourabh Dalvi \IEEEauthorrefmark{2},
        Vincent Carag \IEEEauthorrefmark{2},
        Jiazi Zhang \IEEEauthorrefmark{2}, Jennie Jorgenson\IEEEauthorrefmark{2}, 
        Omar J. Guerra\IEEEauthorrefmark{2}
    }
    \\
    \IEEEauthorblockA{\IEEEauthorrefmark{1} Colorado School of Mines, Advanced Energy Systems Program, Golden, CO}
    \IEEEauthorblockA{\IEEEauthorrefmark{2} National Renewable Energy Laboratory, Golden, CO}
    Email: amoghthatte@mines.edu,
\{sdalvi, vcarag, jzhang, jjorgenson, oguerra\} @nrel.gov}


%


\maketitle

\begin{abstract}

This study addresses the challenges in optimizing long-duration energy storage (LDES) dispatch within future power systems featuring high integration of variable renewable energy (VRE). The research focuses on conducting a comparative analysis between traditional and extended horizon methods for the optimization of LDES dispatch, using open-source and commercial production cost models (PCMs), tested on a futuristic Electric Reliability Council of Texas (ERCOT) grid. The findings indicate that, despite its complexity and longer solution times, the extended horizon approach demonstrates superior performance in LDES dispatch and effectively reduces the impact of degenerate solutions in sequential simulations. This study underscores the trade-offs between computational efficiency and improvement in storage dispatch, which is crucial for future energy systems. The analysis highlights the necessity of addressing the degeneracy issue in storage dispatch in grids dominated by zero operating cost VRE generators and low operating cost energy storage devices. Additionally, the research reveals revenue discrepancies for LDES operators across different models, a consequence of the persistent presence of degeneracy in high VRE systems. These findings suggest an urgent need for refined modeling techniques in the planning and operation of future energy systems.

\end{abstract}


\begin{IEEEkeywords}
High VRE Integration, Unit Commitment, Long Duration Energy Storage, Degeneracy, Production Cost Modeling, PowerSimulations.
\end{IEEEkeywords}

%
\IEEEpeerreviewmaketitle
\vspace{-1.5em}
\section{Introduction}

Electricity grids are anticipated to undergo a transformation with the large number of variable renewable energy (VRE) generators which typically have zero operating cost, complemented by low operating cost long-duration energy storage (LDES) and short-duration energy storage (SDES). Inter-day LDES, characterized by its ability to shift power between multiple days (i.e., 10–36 hours \cite{LDES2023}), and intra-day SDES, typically less than 8 hours, is set to potentially displace a majority of conventional generation in future. However, LDES modeling is a major challenge in power system operations due to the requirement of high temporal resolution  over longer optimization windows to mitigate the mismatch between load and VRE generation spanning multiple time-scales.

Existing generic approaches for modeling LDES include extended optimization horizon, energy targets, and stored energy value. Guerra et al. \cite{Guerra2024} provides a detailed quantitative performance analysis of these methods.
The energy targets approach \cite{Deane2013}, is commonly used in hydro power optimization. This method includes a simplified production cost model (PCM) informed medium-term (MT) simulation stage, where MT phase's storage dispatch is set as state of charge (SOC) targets for standard PCM runs. However, these targets do not capture complex power system dynamics fully, potentially leading to an undervaluation of storage operation \cite{Guerra2024}. 
Stored energy value approach incentivizes storage devices to maintain high SOC by assigning value to the stored energy.
Another approach - time aggregation  \cite{Deml2015} is employed to reduce computational complexity by aggregating non-critical hours. The challenge with both energy value and time-aggregation approaches is in identifying appropriate energy value and critical periods, respectively, which depend on grid conditions.  



Lastly, the extended horizon method \cite{NIET2020} expands the standard one-day optimization window up to a month preventing the premature depletion of energy storage devices by valorizing stored energy over a longer period. While effective in optimizing storage operations considering thermal commitment and network flow constraints, this method significantly increases computation times and memory requirement \cite{Guerra2024}. 
A time horizon that balances computational requirements and operational benefits is subject to the size and type (solar/wind-driven) of system \cite{Guerra2024}
.However, this method's seamless integration into most PCMs makes it suitable for large-scale systems.


The integration of high VRE with both SDES and LDES introduces degeneracy \cite{Lodi2013}, where multiple solutions to a PCM yield the same objective function value. Recent studies have tackled this problem by introducing a time dependent pre-multiplier to the generation commitment variable in the objective function \cite{MARTINEK2018854}, assigning a small cost to storage discharging and to transmission flows \cite{Gates2021}, adding a small randomized variation to all thermal generator properties \cite{FREW20211143}, and incorporating technology-specific constraints \cite{Geth2020}. However, these studies limited their scope to one or more factors including the usage of a capacity expansion model, low VRE integration levels, smaller electricity grid size, and the dominance of a particular type of storage technology operating in the electricity grid.  


This study presents a comparative analysis of traditional and extended horizon approaches for optimizing inter-day LDES dispatch in renewable-dense electricity grids, focusing on a zonal Electric Reliability Council of Texas (ERCOT) system.  The analysis includes a unique comparison between open-source (PowerSimulations) \cite{Lara2023} and commercial (PLEXOS) \cite{PLEXOS} PCM tools, demonstrating how degeneracy can affect results, even when datasets and mathematical representations are identical. The findings are aligned with existing literature, emphasizing that the extended horizon method, though time-intensive, improves the LDES dispatch and mitigates some degeneracy in sequential PCM simulations. The analysis highlights that the traditional approach results in inefficient utilization of LDES, and does not address degeneracy which can lead to \$89M difference in system operation cost. Additionally, the research  indicates the presence of degeneracy in high VRE systems with LDES devices, leading to divergent revenue outcomes for LDES operators, despite closely aligned total system costs.


\vspace{-0.75em}
\section{Methodology}
Traditionally, a PCM is solved with rolling-horizon approach, i.e., 365 problems, each representing a 48-hour operations window offset 24-hours after the previous problem. Each problem minimizes the objective function (\cref{eq:obj}) and is initialized with the results from the previous problem. 


\setlength{\jot}{1pt}
\begin{equation}
\label{eq:obj} 
\begin{gathered}
\mathrm{min} \; \sum_{t} \sum_{i}\left(C_{i}^{f} p_{i,t} + C_{i}^{su} x_{i,t}^{su} + C_{i}^{sd} x_{i,t}^{sd} \right) \\
+ \sum_{t} \sum_{s} \left( C^{op}_{s} p^{d}_{s,t} + C^{op}_{s} p^{c}_{s,t}\right) + \sum_{j} \sum_{t} \mathcal{M}\cdot \mathcal{L}^{\mathrm{drop}}_{j,t} 
\end{gathered}
\end{equation}

\cref{eq:obj} contains parameters $C^{f}_{i}$, $C^{su}_{i}$, $C^{sd}_{i}$ representing fuel, startup, and shutdown costs of thermal generator `$i$', respectively. A nominal operating cost \cite{Cole_2021} ($C^{op}_{s}$= \$0.001/MWh for SDES, \$0.005/MWh for LDES) is assigned to the charging-discharging of storage devices. Based on a common approach observed in the literature \cite{Gates2021,FREW20211143}, the nominal operating cost of storage is varied by introducing a randomized variation of 10\% to avoid degeneracy. $\mathcal{M}$ is a dropped load ($\mathcal{L}^{\mathrm{drop}}$) penalty (\$10,000/MWh). The continuous variables $p_{i,t}$, $p_{vre,t}$, $p^{c}_{s,t}$, $p^{d}_{s,t}$ represent thermal and renewable generation, storage charging and discharging, respectively. While the binary variables $x^{su}_{i,t}$, $x^{sd}_{i,t}$ represent thermal startup and shutdown, respectively. A nodal PCM with reserves contains  the following constraints:

\begin{equation}
\label{eq:dem_sup}
\begin{gathered}
\sum_{i} p_{i,j,t} + \sum_{vre} p_{vre,j,t} + \sum_{s} p^{d}_{s,j,t} + \sum_{k, k\neq j} f_{j,k,t} =  \\
 d_{j,t} + \sum_{s} p^{c}_{s,j,t} + \mathcal{L}^{\mathrm{drop}}_{j,t} \quad \forall j,t
\end{gathered}
\end{equation}

\begin{gather}
P^{\mathrm{min}}_{i,t}x_{i,t} \leq p_{i,t} + \sum_{r} r_{r,i,t} \leq P^{\mathrm{max}}_{i,t}x_{i,t}\quad \forall i,t\label{eq:min_max_th_gen}\\
R^{\mathrm{down}}_{i} \leq p_{i,t} - p_{i,t-1} \leq R^{\mathrm{up}}_{i}\quad \forall i,t\label{eq:ramp_up_down} \\
x_{i,t-1} - x_{i,t} + x^{su}_{i,t} - x^{sd}_{i,t} = 0 \quad \forall i,t\label{eq:track_st_sd} \\
p_{vre,t} + \sum_{r} r_{r,vre,t} \leq P^{\mathrm{Avb}}_{vre,t} \quad \forall vre,t \label{eq:VRE_avb}\\
\sum_{i} r_{r,i,t} + \sum_{vre} r_{r,vre,t} + \sum_{s} r_{r,s,t} \geq RR_{r,t} \quad \forall r,t\label{eq:reserve_prov} \\
0 \leq p^{c}_{s,t} \leq PC_{s,t} x^{c}_{s,t} \quad \forall s,t  \label{eq:st_chg_max} \\
0 \leq p^{d}_{s,t} \leq PD_{s,t} \left(1-x^{c}_{s,t}\right) \quad \forall s,t \label{eq:st_dchg_max} \\
SOC_{s,t} = SOC_{s,t-1} + p^{c}_{s,t}/\eta^{c}_{s} - \eta^{d}_{s}p^{d}_{s,t} \quad \forall s,t \label{eq:track_soc}\\
\eta^{c}_{s\in\mathrm{SDES}} = 0.85, \eta^{c}_{s\in\mathrm{LDES}} = 0.65  \label{eq:SDES_eff}\\
\eta^{d}_{s\in\mathrm{SDES}} = \eta^{d}_{s\in\mathrm{LDES}} = 1 \label{eq:LDES_eff}\\
SOC^{\mathrm{min}}_{s} \leq SOC_{s,t} \leq SOC^{\mathrm{max}}_{s} \quad \forall s,t \label{eq:soc_limit}\\
SOC^{\mathrm{max}}_{s} = 4 \cdot PD_{s} \quad \forall s \in \mathrm{SDES}\label{eq:SDES_storage_duration}\\
SOC^{\mathrm{max}}_{s} = 10 \cdot PD_{s} \quad \forall s \in \mathrm{LDES}\label{eq:LDES_storage_duration}\\
SOC^{\mathrm{min}}_{s} = 0.1 \cdot SOC^{\mathrm{max}}_{s} \quad \forall s \label{eq:min_storage}\\
SOC_{s,t=1} = 0.5 \cdot SOC^{\mathrm{max}}_{s} \quad \forall s \label{eq:initial_storage}\\
p^{d}_{s,t} + \sum_{r} r_{r,s,t} \leq PD_{s} + p^{c}_{s,t} \quad \forall s,t \label{eq:st_res_prov} \\
F^{\mathrm{min}}_{j,k} \leq f_{j,k,t} \leq F^{\mathrm{max}}_{j,k} \quad \forall j,k,t \label{eq:line_limit}\\
f_{j,k,t} = B_{j,k}\left(\theta_{j,t} - \theta_{k,t} \right) \quad \forall j,k,t \label{eq:line_angle_calc}\\
\Theta^{\mathrm{min}} \leq \theta_{j,t} \leq \Theta^{\mathrm{max}} \quad \forall j,t.\label{eq:angle_limit}
\end{gather}

\cref{eq:dem_sup} is a demand-supply balance for all nodes `$j$' and time `$t$', where $f_{j,k,t}$ and $d_{j,t}$ is power line flow and demand, respectively. \cref{eq:ramp_up_down} specifies thermal generator ramping limits with the ramp down ($R^{\mathrm{down}}_{i}$) and ramp up ($R^{\mathrm{up}}_{i}$) parameters. \cref{eq:track_st_sd} tracks the thermal generator status, where $x_{i,t}$ represents a binary ON/OFF variable for a thermal generator. \cref{eq:VRE_avb} limits VRE generation through VRE availability ($P^{\mathrm{Avb}}_{vre,t}$). \cref{eq:reserve_prov} ensures the provision of specified reserve type `$r$' ($RR_{r,t}$) by thermal generators ($r_{r,i,t}$) and storage devices ($r_{r,s,t}$). Eqs.~\ref{eq:st_chg_max}-\ref{eq:st_dchg_max} provide maximum charging ($PC_{s}$) and discharging ($PD_{s}$) limits for the storage, respectively. The binary variable $x^{c}_{s,t}$ ensures that the storage operates in one mode at a time. \cref{eq:track_soc} book keeps the state of charge ($SOC_{s,t}$) in a storage with the help of storage charging ($\eta^{c}_{s}$) and discharging ($\eta^{d}_{s}$) efficiencies, while \cref{eq:soc_limit} provides bounds to the SOC. \cref{eq:st_res_prov} represents raise reserve services provided by energy storage devices, by either increasing the power discharge or decreasing the charging power at a given time. Note that \cref{eq:SDES_eff,eq:LDES_eff,eq:SDES_storage_duration,eq:LDES_storage_duration,eq:min_storage,eq:initial_storage} are storage-related study-specific constraints.  \cref{eq:line_limit} provides bounds to the transmission line flows. \cref{eq:line_angle_calc} evaluates bus voltage angles ($\theta_{j,t}$) with the help of line susceptance ($B_{j,k}$) and \cref{eq:angle_limit} specifies limits for the bus voltage angle. 

This study adopted a 7-day look-ahead  (representing a 192-hour operations window offset 24-hours after the previous problem), as longer horizons  yield diminishing improvements in total operating costs while significantly increasing solve times \cite{Guerra2024}. By integrating daily unit commitment constraints, the model optimizes stored energy to align with anticipated energy demands. The additional look-ahead period serves the dual purpose of mitigating myopic decision making and constricting the solution space - reducing number of degenerate solutions. However, a rigorous mathematical proof of reduction in degeneracy is beyond the scope of this work.



\vspace{-0.5em}
\section{Metrics for Comparison}


Energy storage is characterised by equivalent storage cycles \cite{Guerra_2021} that provides a metric to understand the extent of storage cycling in an electricity grid. The cumulative difference in the total production cost (TPC) ($\Delta TPC$) is useful to quantify how the difference between the TPCs of two models grow with time. It is defined as, 



\begin{equation}
\Delta TPC_{d} = \sum_{d} \left(TPC^{\mathrm{PL}}_{d} - TPC^{\mathrm{PS}}_{d} \right).\label{eq:cum_diff_tpc}
\end{equation}

In the above equation, $TPC^{\mathrm{PL}}_{d}$ and $TPC^{\mathrm{PS}}_{d}$ denote TPC on a day `$d$' predicted by PLEXOS and PowerSimulations models, respectively. The difference in the SOC ($\Delta SOC_{\mathrm{tech},t}$ - expressed as a percentage) is given by, 

\begin{equation}
\Delta SOC_{\mathrm{tech},t} = \frac{\sum_{s \in \mathrm{tech}}\left(SOC^{\mathrm{PS}}_{s,t} - SOC^{\mathrm{PL}}_{s,t}\right)}{\sum_{s \in \mathrm{tech}} CapES_{s}}.\label{eq:soc_diff}
\end{equation}


The storage technology's revenue (\cref{eq:revenue_storage}) is calculated with the help of locational marginal price ($\mathrm{LMP}_{t}$).  

\begin{equation}
\mathrm{Revenue}_{tech} = \sum_{s \in \mathrm{tech}} \sum_{t \in NT} \left(\mathrm{LMP}_{t} \left( p^{d}_{s,t} - p^{c}_{s,t}\right) \right)\label{eq:revenue_storage}
\end{equation}




\vspace{-0.5em}
\section{System Description}
The methodology tested the futuristic ERCOT system on two PCM platforms, PLEXOS and PowerSimulations that use Gurobi \cite{gurobi} and Xpress \cite{xpress2014fico} solvers, respectively. The ERCOT grid is carved out of the ReEDS system \cite{ReEDS_2021} and is represented by 7 balancing areas interconnected with 12 transmission lines. The renewable energy potential (reV) model \cite{ReV_2019} provided the hourly timeseries for the VRE generators. 
Detailed information on the installed generation and battery capacities is now publicly available \cite{Dalvi2024}.
The PowerSimulations simulations uses dual socket Intel Xeon Sapphire Rapids (52-core) processors, with 256 GB DDR5 memory while the PLEXOS uses Intel(R) Xenon(R) Gold 6330 CPU (14-core) processors with 256 GB memory. Simulations are run on two different machines due to licensing limitation of commercial software. The relative MIP gap is set at 1E-4 and the time limit is 1000 seconds. 
Maximum number of days in both traditional (324 days) and extended horizon (321 days) were solved within the set time limit with given MIP gap, showing 1000s is sufficient solve time.
A typical summer week with fluctuating daily net load is selected for bench marking two models .

\vspace{-0.5em}
\section{Results}
The models benchmarking shows identical model inputs and outputs-TPCs for the unit commitment (UC) problem with 168 hours optimization horizon. Table \ref{tab:validation_effors} shows that the relative difference (calculated with respect to PLEXOS) between the total cost of two models is 9.3E-7 (smaller than the relative MIP gap). Both models predict identical VRE curtailment, generation by the individual thermal generators, and storage discharging. The difference in wind and solar generation stems from the degeneracy associated with zero operating cost VRE generation. Since the models indicated initial agreement, the next phase involved running two models for one year with the traditional (1-day ahead) and extended horizon approach (7-day ahead) using the inputs and settings. 

\begin{table}[!h]
\centering
\caption{Validating PowerSimulations and PLEXOS models with 168 hours optimization (May 7- 13) run}
\label{tab:validation_effors}
\begin{tabular}{@{}lrr@{}}
\toprule
Modeling platform            & \multicolumn{1}{c}{PowerSimulations} & \multicolumn{1}{c}{PLEXOS} \\ \midrule
Total production cost {[}\${]}          & 7,499,790                &  7,499,720          \\
VRE curtailment {[}GWh{]}    & 487                          & 487                          \\
\arrayrulecolor{black!30}\midrule
Thermal generation {[}GWh{]} & 16                       & 16                       \\
Nuclear generation {[}GWh{]} & 833                        & 833                        \\
Solar generation {[}GWh{]}      & 2939                       & 2975                       \\
Wind generation {[}GWh{]}    & 4198                       & 4162                       \\
\arrayrulecolor{black!30}\midrule
SDES discharge {[}GWh{]}     & 438                        & 438                        \\
LDES discharge {[}GWh{]}     & 85                         & 85                         \\
 \arrayrulecolor{black}\bottomrule
\end{tabular}
\end{table}

The effect of degeneracy becomes visible for 1-year solution solved with traditional approach. Table \ref{tab:1_year_run_comparison} highlights an increase in the difference between the TPC: \$89M (relative gap- 7E-4) and \$3M (relative gap- 2E-5) for traditional and extended horizon runs, respectively. The most distinguishing feature is that even though the difference between the TPCs two models is on the order of the relative MIP gap, the storage operation differs significantly. 


\begin{figure}[!h]
\centering
\includegraphics[width=3in]{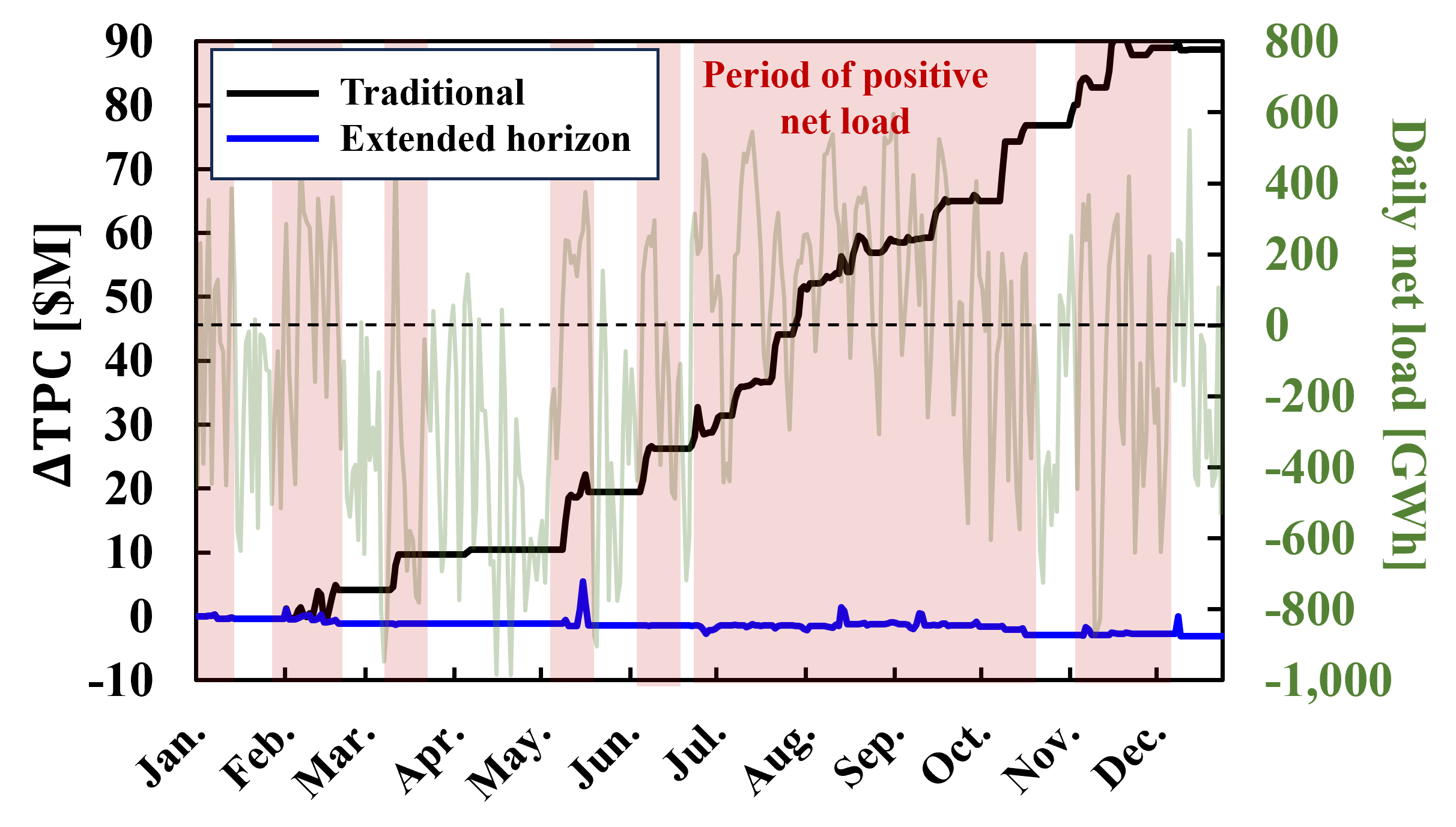}
\caption{Comparison of the evolution of the cumulative difference in the production cost for traditional and extended horizon simulations plotted with the daily net-load}
\vspace{-0.3em}
\label{fig:cum_prod_cost}
\end{figure}


The difference in the storage operation affects the VRE and thermal generation, ultimately affecting the TPC. Table \ref{tab:1_year_run_comparison} shows that the PowerSimulations storage dispatch for LDES and SDES is higher by 3020 GWh and 2731 GWh, respectively for the traditional simulation. Increase in the storage dispatch is supported by more charging of the storage devices predominantly through the VRE generation, which is higher by 2620 GWh in PowerSimulations. As a result, VRE curtailment is smaller in PowerSimulations by 0.5\%. On the other hand, lower storage dispatch in the PLEXOS model causes increase in its thermal generation by 511 GWh, thus increasing its TPC.

\begin{figure}[!h]
\centering
\includegraphics[width=2.5in]{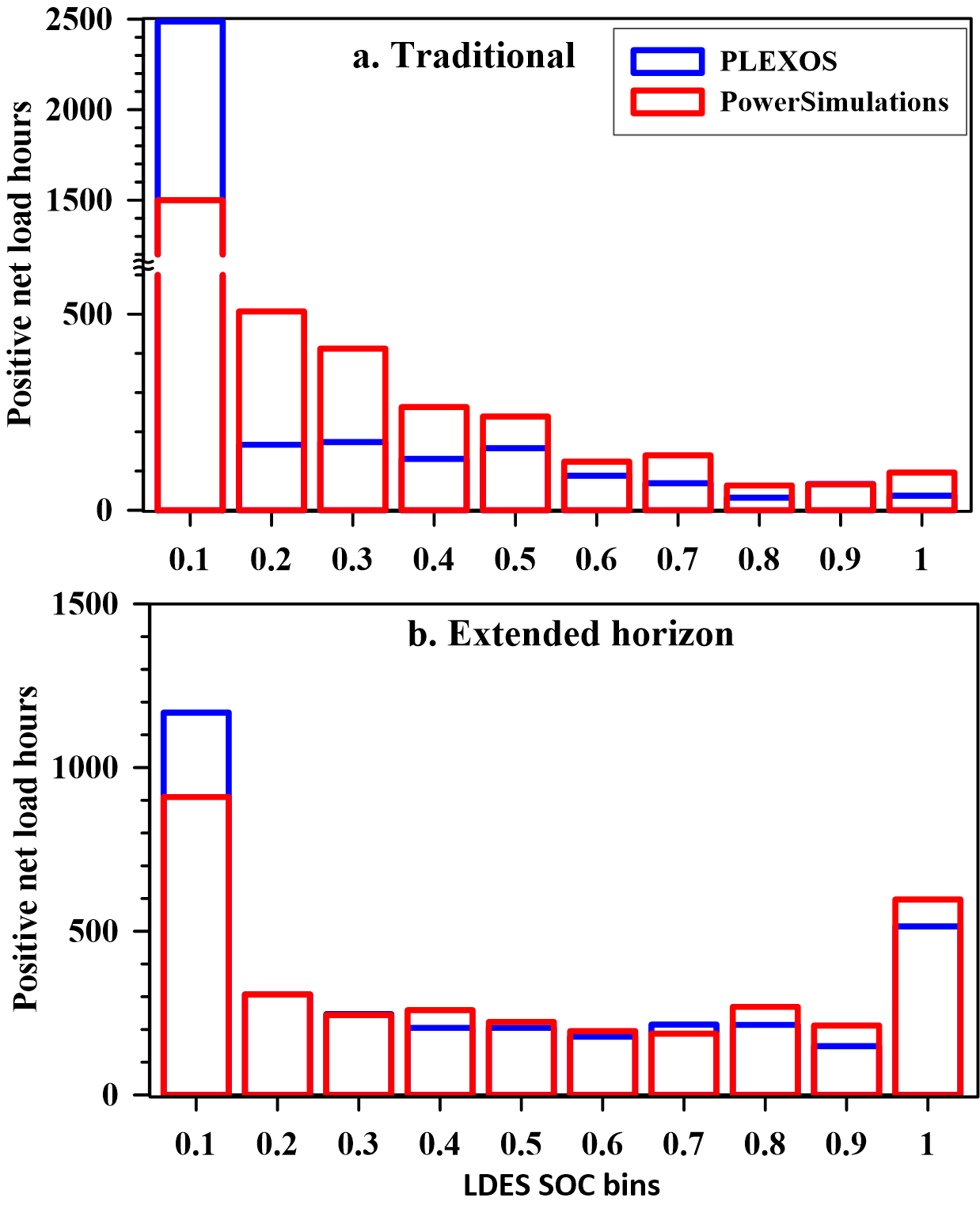}
\caption{LDES SOC histogram plot for the traditional and extended horizon PLEXOS and PowerSimulations models}
\label{fig:SOC_diff}
\end{figure}

The difference in the TPC arises from the accumulation of higher costs during the positive net-load days (periods highlighted by red color in Fig.~\ref{fig:cum_prod_cost}). This indicates that the PowerSimulations (Xpress solver) is  opting for a solution that reduces curtailment to charge storages more during the negative-net load days, which displaces conventional generation during the positive net-load days. Fig.~\ref{fig:SOC_diff}a shows that LDES in PowerSimulations is maitaining higher SOC for more number of positive net load hours reducing the net thermal generation and hence the TPC for PowerSimulations. A five fold increase in the number of positive net load hours during which both models are maintaining 100\% SOC in the LDES as compared to the traditional case (Fig.~\ref{fig:SOC_diff}b) demonstrates the potential of extended horizon method in improving the LDES operation.



\begin{figure*}[!ht]
\centering
\includegraphics[width=6.6in]{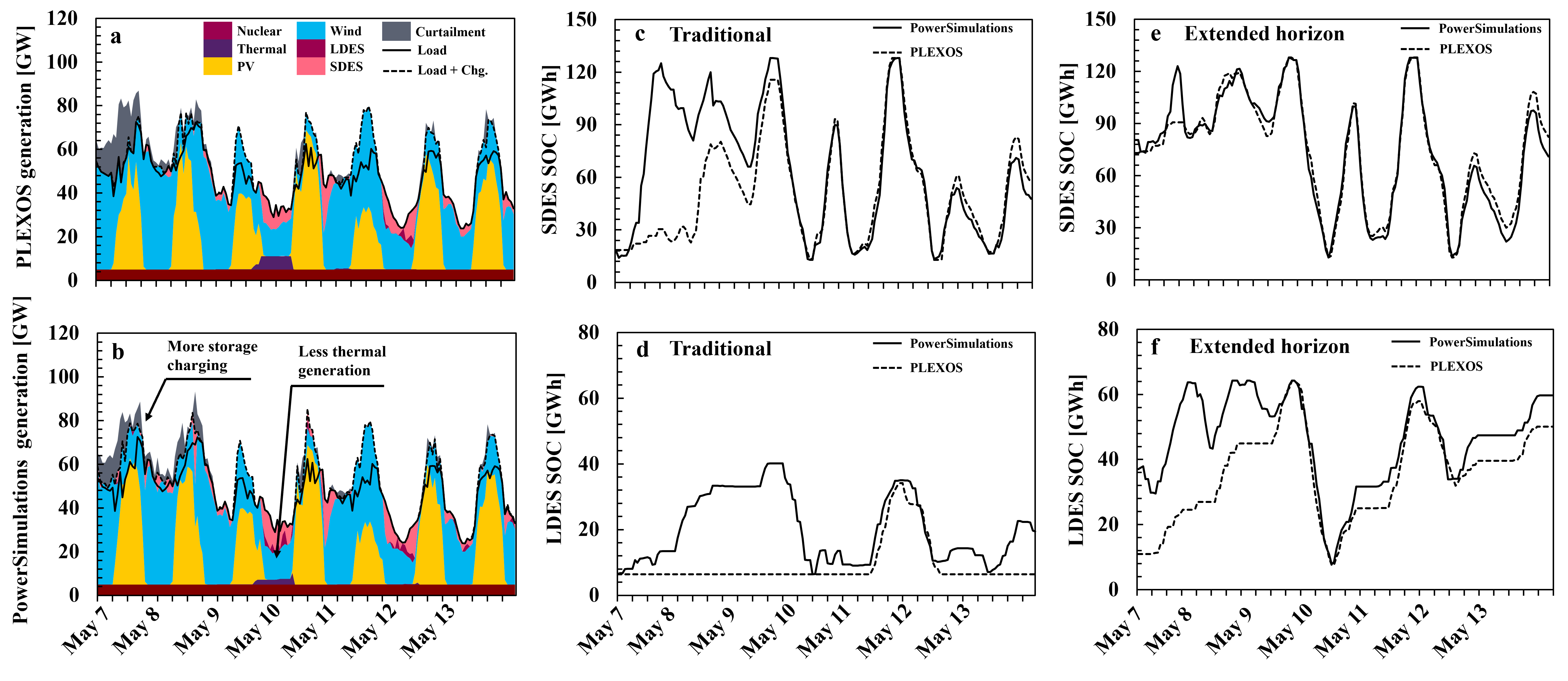}
\caption{Generation dispatch stacks (a,b) for the traditional approach and storage SOCs for the traditional (c,d) and extended horizon approaches (e,f) with PLEXOS and PowerSimulations models during May 7-13}
\vspace{-0.5em}
\label{fig:crit_week}
\end{figure*}

Figs.~\ref{fig:crit_week}a and b show the dispatch stack for the two models during May 7-13, a week characterised by fluctuating daily net load.The TPC predicted by both models for the first three days is almost identical (except negligible differences caused by the storage operations) as nuclear are the only significant cost-contributing generators operating. Interestingly, even though both storage devices begin the week with almost identical SOC, the degeneracy associated with zero-cost VRE and low-cost storage operations allows for the selection of different storage operations, causing PowerSimulations to store   
\begin{table*}[!ht]
\centering
\caption{Comparing annual PLEXOS and PowerSimulations results for different look-ahead periods}
\label{tab:1_year_run_comparison}
\begin{tabular}{@{}lrrrr@{}}
\toprule
& \multicolumn{2}{c}{Traditional}                      & \multicolumn{2}{c}{Extended horizon}                         \\ 
& \multicolumn{1}{l}{PowerSimulations} & \multicolumn{1}{l}{PLEXOS} & \multicolumn{1}{l}{PowerSimulations} & \multicolumn{1}{l}{PLEXOS} \\
\midrule
Total production cost {[}\$M{]} & 3,320                      & 3,409                      & 3,080                      & 3,077                      \\
\arrayrulecolor{black!30}\midrule
Thermal generation {[}TWh{]}            & 60.3  & 60.8  & 58.3 & 58.3  \\       
Solar generation {[}TWh{]}                 & 142.6 & 132.2 & 143.1  & 132.0  \\
Wind generation {[}TWh{]}               & 270.5 & 278.2 & 264.4 & 273.5  \\
SDES discharge {[}TWh{]}                 & 16.6  & 13.9  & 16.9 & 13.4     \\
LDES discharge {[}TWh{]}                 & 4.3 & 1.3 & 4.5 & 2.0 \\
\arrayrulecolor{black!30}\midrule
SDES charge {[}TWh{]}                 & 
19.5 & 16.3 & 19.9 & 15.8 \\
LDES charge {[}TWh{]}                   &  6.6  & 2.0 & 6.9 & 3.1    \\
\arrayrulecolor{black!30}\midrule
Annual SDES eqv. cycles          & 199.5 & 188.8 & 194.9 & 181.9  \\     
Annual LDES eqv. cycles             & 73.9  & 26.2  &  76.4 & 34.9 \\
\arrayrulecolor{black!30}\midrule
VRE curtailment {[}\%{]}                & 20.9         & 21.4                       & 20.8      & 21.2\\
\arrayrulecolor{black!30}\midrule
{Average LMP {[}\$/MWh{]}} & {104.8} & {169.3} & {102.5} & {190.0} \\
{Standard deviation in LMP {[}\$/MWh{]}} & {79.3} & {80.7} & {78.1} & {78.8} \\
{SDES revenue: energy arbitrage {[}\$M/MW-year{]}} & {15.1} & {13.9} &
  {20.4} & {21.2} \\
{LDES revenue: energy arbitrage {[}\$M/MW-year{]}} & {31.6} & {21.4} &
  {51.6} & {49.9} \\  
\arrayrulecolor{black!30}\midrule
{Compute time {[}hrs.{]}} & 3.64 & 17.90 & 13.94 &  32.33\\  
\arrayrulecolor{black}\bottomrule
\end{tabular}
\end{table*}

significantly more energy (Figs.~\ref{fig:crit_week}c,d). As a result, the peaker thermal generation during this week is higher in PLEXOS by 49 GWh causing an increase in the TPC by \$8.5M.

Solving a degenerate UC problem with the traditional approach affects the initial conditions for the next day's simulation, because the end of the day storage SOC and thermal generators ON/OFF status can differ in two models. This transforms the same UC problem into two with different initial conditions on a subsequent day. As a result, the cumulative difference between the objective functions of the two models continues to widen with an increasing number of steps (as seen in Fig.~\ref{fig:cum_prod_cost}). One way to reduce degeneracy is by adopting an extended horizon approach., in which both models can access more information due about future days. Table \ref{tab:1_year_run_comparison} shows that the relative difference between the total thermal generation cost (2E-5) is smaller than the set relative MIP gap (1E-4) for a extended horizon run. However, this approach is not sufficient enough to mitigate the difference in the storage operation as both LDES and SDES storage devices dispatch significantly more (2422 GWh and 3544 GWh, respectively) in PowerSimulations. Solving degeneracy needs further advancements on numerical front \cite{GAMRATH2020,Lodi2013}. The extended horizon approach has limitations on being computationally efficient or scalable.        

The effects of degeneracy becomes more relevant from the perspective of a storage device operator. Table \ref{tab:1_year_run_comparison} shows that the annual energy arbitrage revenues are considerably different for different storage devices even for a extended horizon run. It is important to note that this work does not simulate bidding behavior for the storages, making them effectively zero operating cost, which makes expensive thermal units as price setting units. Therefore, the PLEXOS model featuring high energy prices due to more thermal generation also has relatively higher standard deviation in energy prices.    



\vspace{-0.5em}
\section{Conclusion}


This work highlights the trade-offs between computational efficiency and improvement in LDES dispatch, crucial for high VRE integrated energy systems with LDES presence. The findings indicate that, despite its complexity and longer solution times, the extended horizon approach demonstrates superior performance in LDES dispatch and effectively reduces the impact of degenerate solutions in sequential simulations. However, the results also suggest that while extending the optimization horizon can mitigate some discrepancies, it does not fully resolve the issue and may introduce computational inefficiencies. 

The revelation of significant differences in storage operation costs and VRE curtailment between two PCMs underlines the impact of degeneracy on TPCs and the operation of storage devices in high VRE scenarios. The observed disparity in energy arbitrage revenues for LDES operators across different models, arising from the presence of degeneracy in high VRE systems, opens up several avenues for future research. These findings point towards the urgent need for refined modeling techniques in the planning and operation of future energy systems. Consequently, future research should focus on developing more sophisticated LDES dispatch strategies, examining the applicability of diverse modeling approaches in various grid scenarios, and supporting the transition towards more sustainable and reliable energy systems.



%


\vspace{-0.5em}
\section*{Acknowledgment}

The authors would like to thank Paul Denholm for helpful discussions that took the research forward. This work was authored by the National Renewable Energy Laboratory, operated by Alliance for Sustainable Energy, LLC,
for the U.S. Department of Energy (DOE) under Contract No. DE-EE00038429. The U.S. Government retains and the publisher,by accepting the article for publication, acknowledges that the U.S. Government retains a nonexclusive, paid-up, irrevocable, worldwide license to publish or reproduce the published form of this work, or allow others to do so, for U.S. Government purposes.

\ifCLASSOPTIONcaptionsoff
  \newpage
\fi



\bibliographystyle{IEEEtran}
\bibliography{IEEEabrv,pes_conf23.bib}

\begin{thebibliography}{10}
\providecommand{\url}[1]{#1}
\csname url@samestyle\endcsname
\providecommand{\newblock}{\relax}
\providecommand{\bibinfo}[2]{#2}
\providecommand{\BIBentrySTDinterwordspacing}{\spaceskip=0pt\relax}
\providecommand{\BIBentryALTinterwordstretchfactor}{4}
\providecommand{\BIBentryALTinterwordspacing}{\spaceskip=\fontdimen2\font plus
\BIBentryALTinterwordstretchfactor\fontdimen3\font minus \fontdimen4\font\relax}
\providecommand{\BIBforeignlanguage}[2]{{%
\expandafter\ifx\csname l@#1\endcsname\relax
\typeout{** WARNING: IEEEtran.bst: No hyphenation pattern has been}%
\typeout{** loaded for the language `#1'. Using the pattern for}%
\typeout{** the default language instead.}%
\else
\language=\csname l@#1\endcsname
\fi
#2}}
\providecommand{\BIBdecl}{\relax}
\BIBdecl

\bibitem{LDES2023}
\BIBentryALTinterwordspacing
{US DOE}. {The Pathway to Long Duration Energy Storage: Commercial Liftoff}. [Online]. Available: \url{https://liftoff.energy.gov/long-duration-energy-storage/}
\BIBentrySTDinterwordspacing

\bibitem{Guerra2024}
\BIBentryALTinterwordspacing
O.~J. Guerra, S.~Dalvi, A.~A. Thatte, B.~Cowiestoll, J.~Jorgenson, , and B.~Hodge, ``{Towards Robust and Scalable Dispatch Modeling of Long-Duration Energy Storage},'' \emph{arXiv preprint}, 2024. [Online]. Available: \url{https://arxiv.org/abs/2401.16605}
\BIBentrySTDinterwordspacing

\bibitem{Deane2013}
J.~P. Deane, E.~J. McKeogh, and B.~P.~O. Gallachoir, ``Derivation of intertemporal targets for large pumped hydro energy storage with stochastic optimization,'' \emph{IEEE Transactions on Power Systems}, vol.~28, no.~3, pp. 2147--2155, 2013, doi:{10.1109/TPWRS.2012.2236111}.

\bibitem{Deml2015}
S.~Deml, A.~Ulbig, T.~Borsche, and G.~Andersson, ``The role of aggregation in power system simulation,'' in \emph{2015 IEEE Eindhoven PowerTech}, 2015, pp. 1--6, doi:{10.1109/PTC.2015.7232755}.

\bibitem{NIET2020}
T.~Niet, ``Storage end effects: An evaluation of common storage modelling assumptions,'' \emph{Journal of Energy Storage}, vol.~27, p. 101050, 2020, doi:{10.1016/j.est.2019.101050}.

\bibitem{Lodi2013}
A.~Lodi and A.~Tramontani, \emph{Performance Variability in Mixed-Integer Programming}, ch. Chapter 1, pp. 1--12, doi:{10.1287/educ.2013.0112}.

\bibitem{MARTINEK2018854}
J.~Martinek, J.~Jorgenson, M.~Mehos, and P.~Denholm, ``A comparison of price-taker and production cost models for determining system value, revenue, and scheduling of concentrating solar power plants,'' \emph{Applied Energy}, vol. 231, pp. 854--865, 2018, doi:{10.1016/j.apenergy.2018.09.136}.

\bibitem{Gates2021}
N.~Gates, W.~Cole, A.~W. Frazier, and P.~Gagnon, ``Evaluating the {I}nteractions {B}etween {V}ariable {R}enewable {E}nergy and {D}iurnal {S}torage,'' NREL, Tech. Rep., Oct. 2021, doi:{10.2172/1827634}.

\bibitem{FREW20211143}
B.~Frew, B.~Sergi, P.~Denholm, W.~Cole, N.~Gates, D.~Levie, and R.~Margolis, ``The curtailment paradox in the transition to high solar power systems,'' \emph{Joule}, vol.~5, no.~5, pp. 1143--1167, 2021, doi:{10.1016/j.joule.2021.03.021}.

\bibitem{Geth2020}
F.~Geth, C.~Coffrin, and D.~Fobes, ``A flexible storage model for power network optimization,'' in \emph{Proceedings of the Eleventh ACM International Conference on Future Energy Systems}, 06 2020, doi:{10.1145/3396851.3402121}.

\bibitem{Lara2023}
J.~D. Lara, S.~Dalvi, C.~Barrows, D.~Thom, Lilyhanig, R.~Henríquez-Auba, D.~Krishnamurthy, P.~Monticone, R.~Saavedra, M.~Irish, dsigler1234, O.~Dowson, J.~Maack, M.~Kratochvil, J.~TagBot, M.~Cawte, T.~Borbáth, T.~G. Badger, T.~Holy, alefcastelli, C.~Coffrin, and timkittel, ``{NREL-Sienna/PowerSimulations.jl: v0.24.1},'' Nov. 2023, doi:{10.5281/zenodo.10060186}.

\bibitem{PLEXOS}
{Energy Exemplar}, ``{PLEXOS}.''

\bibitem{Cole_2021}
W.~Cole, A.~W. Frazier, and C.~Augustine, ``Cost projections for utility-scale battery storage: 2021 update,'' NREL, Tech. Rep., Jun. 2021, doi:{10.2172/1786976}.

\bibitem{Guerra_2021}
O.~J. Guerra, J.~Eichman, and P.~Denholm, ``Optimal energy storage portfolio for high and ultrahigh carbon-free and renewable power systems,'' \emph{Energy Environ. Sci.}, vol.~14, pp. 5132--5146, 2021, doi:{"10.1039/D1EE01835C"}.

\bibitem{gurobi}
{Gurobi Optimizer Reference Manual}.

\bibitem{xpress2014fico}
{FICO Xpress Optimization Suite}.

\bibitem{ReEDS_2021}
J.~Ho, J.~Becker, M.~Brown, P.~Brown, I.~Chernyakhovskiy, S.~Cohen, W.~Cole, S.~Corcoran, K.~Eurek, W.~Frazier, P.~Gagnon, N.~Gates, D.~Greer, P.~Jadun, S.~Khanal, S.~Machen, M.~Macmillan, T.~Mai, M.~Mowers, C.~Murphy, A.~Rose, A.~Schleifer, B.~Sergi, D.~Steinberg, Y.~Sun, and E.~Zhou, ``Regional {E}nergy {D}eployment {S}ystem ({ReEDS}) {M}odel {D}ocumentation ({V}ersion 2020),'' NREL, Tech. Rep., Jun. 2021, doi:{10.2172/1788425}.

\bibitem{ReV_2019}
G.~Maclaurin, N.~Grue, A.~Lopez, D.~Heimiller, M.~Rossol, G.~Buster, and T.~Williams, ``The {R}enewable {E}nergy {P}otential {(reV)} {M}odel: {A} {G}eospatial {P}latform for {T}echnical {P}otential and {S}upply {C}urve {M}odeling,'' NREL, Tech. Rep., Sep. 2019, doi:{10.2172/1563140}.

\bibitem{Dalvi2024}
\BIBentryALTinterwordspacing
S.~Dalvi. (2024) {ERCOT-Energy-Storage-Study-Dataset}. GitHub. [Online]. Available: \url{https://github.com/NREL/ERCOT-Energy-Storage-Study-Dataset.git}
\BIBentrySTDinterwordspacing

\bibitem{GAMRATH2020}
G.~Gamrath, T.~Berthold, and D.~Salvagnin, ``An exploratory computational analysis of dual degeneracy in mixed-integer programming,'' \emph{{EURO Journal on Computational Optimization}}, vol.~8, no.~3, pp. 241--261, 2020, doi:{10.1007/s13675-020-00130-z}.

\end{thebibliography}
%



%
\vspace{-0.5em}
\begin{IEEEbiography}[{\includegraphics[width=1in,height=1.25in,clip,keepaspectratio]{picture}}]{John Doe}
\blindtext
\end{IEEEbiography}




\end{document}